\documentclass[prl,twocolumn]{revtex4-1}

\usepackage{graphicx}
\usepackage{amsmath}
\usepackage{threeparttable}
\usepackage{natbib}

\usepackage{lineno}

\usepackage{threeparttable}

\begin{document}


\title{Four hints and test candidates of the local cosmic expansion} 


\author{Kazuhiro Agatsuma}
\email[]{agatsuma@star.sr.bham.ac.uk}
\affiliation{School of Physics and Astronomy and Institute for Gravitational Wave Astronomy, University of Birmingham, Edgbaston, Birmingham B15 2TT, United Kingdom}


\date{\today}

\begin{abstract}
The expansion of the universe on short distance scales is a new frontier to investigate the dark energy. 
The excess orbital decay in binary pulsars may be related to acceleration by the local cosmic expansion, called the cosmic drag. 
Modern observations of two independent binaries (PSR B1534+12 and PSR B1913+16) support this interpretation 
and result in a scale-independent expansion with viscous uniformity, 
in which binary systems have a smaller expansion rate than the Hubble constant. 
This paper shows additional evidential binaries (PSR J1012--5307 and PSR J1906+0746), supporting the cosmic drag picture. 
The total anomaly of the conventional model is about $3.6\,\sigma$ including two evidential binaries reported before. 
In addition, an observable range of the cosmic drag has been calculated for typical models of both NS-NS binary and NS-WD binary. 
In this region, six test candidates are listed with predictions of the excess orbital decay. 
\end{abstract}


\maketitle

\section{Introduction}
A discovery of the expansion of the universe made a paradigm shift. 
The recession velocities of distant galaxies follow the Hubble--Lema\^{i}tre law~\cite{Hubble,Lemaitre}. 
In these two decades, progress in observations revealed that the cosmic expansion is accelerating~\cite{Riess_1998,Perlmutter_1999}. 
The acceleration mechanism, which is called the dark energy, is one of the most intriguing mysteries in our universe. 
The cosmic expansion is visible on large scale observations but not valid on small scales ($\lesssim 70$\,Mpc)  
because local peculiar velocities exceed the recession velocity~\cite{Freedman_2001}. 
Since the gravitationally bound systems maintain own size, 
the cosmic expansion is separated from local dynamics as the current standard interpretation. 

Another interpretation survives with following general relativity (GR)~\cite{Anderson1995,Cooperstock1998,Bonnor1999,Adkins2007,Carrera2010,Price2012,Giulini2014}. 
They argue that the cosmic expansion can be acceleration on local dynamics. 
The formulation was built using the geodesic deviation equation~\cite{Cooperstock1998}, 
in which Friedmann-Lema\^{i}tre-Robertson-Walker (FLRW) coordinates is transformed to Fermi normal coordinates 
as an approximation of the locally inertial frame. 
The equations of motion to the lowest order are expressed by 
\begin{equation}
 \frac { d ^ { 2 } x _ { \mathrm { F } } ^ { k } } { d t ^ { 2 } } - \left( \frac { \ddot { a } } { a } \right) x _ { \mathrm { F } } ^ { k } = 0
\label{eq:FRW}
\end{equation}
where 
$ x _ { \mathrm { F } } ^ { k } $ is the geodesic distance in the Fermi normal coordinates with $k$ from 1 to 3,
and $ a $ is the scale factor. 
The expansion term is called the cosmic drag~\cite{Anderson1995}, 
which represents a stretch of space. 
The object's position is not fixed on space but dragged by a flow of the expanding space.  
In this framework, the Hubble--Lema\^{i}tre law is related to the cosmic drag by 
\begin{equation}
 \ddot{R}_\mathrm{H} = \dot{H} R + H \dot{R} = (\ddot{a}/a)R = -q H^{2} R . 
\label{eq:RHacc}
\end{equation}
The last term employs the deceleration parameter $q$, 
which is defined as 
\begin{equation}
 q := -\ddot{a}a/\dot{a}^2 = -(\ddot{a}/\dot{a})H^{-1} = -(\ddot{a}/a)H^{-2}
\label{eq:q}
\end{equation}
using $H=\dot{a}/a$ and $\dot{H} = (\ddot{a}/a) - H^2$. 
This expression has been applied to orbital systems~\cite{Carrera2010,Giulini2014}. 
The two body problem is expressed by the pseudo-Newtonian approach: 
\begin{equation}
 \ddot{R} = L^{2}/R^{3} - GM_0/R^{2} -q H^{2} R  
\label{eq:impEOM1}
\end{equation}
with $ L = R^{2} \dot{\phi}$. 
Here, 
$ R $ is the radial geodesic distance, 
$ L $ the conserved angular momentum per unit mass in planar polar coordinates ($R, \phi$), 
$ G $ the gravitational constant, 
$M_0$ the mass of the central body, 
and the present value of $H$ is the Hubble constant $ H_0 \approx 70 $\,km $\mathrm{s}^{-1} \mathrm{Mpc}^{-1}$ 
(this value is used in this paper). 
The same expression can be applied for the scales of galaxy clusters~\cite{Nandra_1,Nandra_2}. 
A critical radius can exist at 
\begin{equation}
R_\mathrm{c} = \left( \frac{GM_0}{- q_0 H_{0}^{2}} \right)^{1/3} 
\label{eq:Rc1}
\end{equation}
for $ q_0 < 0 $ as the present cosmological value of $q$, 
in which the gravitational attraction force equals the local expansion ($ L=0 $ and $\ddot{R}=0$). 
This model is consistent with observations 
because the recession velocity asymptotes the Hubble--Lema\^{i}tre law ($\dot{R}=H_0 R $) on large scales ($R \gg R_\mathrm{c}$). 
However, we cannot test this theory by searching for deviations from standard Keplerian orbits 
because they are too small to measure~\cite{Carrera2010,Giulini2014}. 

A unique approach has been proposed to observe this local expansion effect using binary pulsar systems~\cite{Agatsuma_2020}. 
By comparing with accurate predictions of the orbital decay by gravitational wave (GW) emission, 
it is shown that the braking effect of the local cosmic expansion can be observed by precise measurements. 
The orbital decay of a binary system is given by 
\begin{eqnarray}
\frac{dr}{dt} 
 &=& H_\mathrm{s} \cdot r - \frac{64}{5} \frac{G^3}{c^5} \frac{(m_\mathrm{1} m_\mathrm{2}) (m_\mathrm{1} + m_\mathrm{2}) }{r^3} \notag \\
 &\equiv& H_\mathrm{s} \cdot r - \frac{K_\mathrm{gw}}{r^3} ,
\label{eq:drdt}
\end{eqnarray}
including both the local expansion and GW emission~\cite{Agatsuma_2020,PetersMathews,Peters}. 
Here, 
$ c $ is the speed of light, 
$ r $ is the orbital radius (separation) of binary stars, 
$ m_1 $ and $ m_2 $ are each mass of the binary, 
and $ H_\mathrm{s} $ is the Hubble parameter in this spiral decay system. 
By differentiating Eq.\,(\ref{eq:drdt}) and using Eq.\,(\ref{eq:RHacc}), effective force expression is obtained: 
\begin{equation}
\ddot{r} = - q_\mathrm{s} H_\mathrm{s}^{2} r  - 3 K_\mathrm{gw}^2 r^{-7} . 
\label{eq:impEMGW}
\end{equation}
A purely Keplerian orbit is achieved at the critical radius $r_\mathrm{c}$ where the local expansion equals GW reaction. 
At this point, $ dr/dt = 0 $ and $\ddot{r}=0$ are imposed in Eq.\,(\ref{eq:drdt}) and Eq.\,(\ref{eq:impEMGW}), respectively. 
These provide 
\begin{equation}
r_\mathrm{c} 
= \left( \frac{ K_\mathrm{gw} }{ H_\mathrm{s} }  \right)^{1/4} 
\label{eq:Rmax}
\end{equation}
and 
\begin{equation}
r_\mathrm{c} 
= \left( \sqrt{ - \frac{3}{q_\mathrm{s}}} \frac{ K_\mathrm{gw} }{ H_\mathrm{s} } \right)^{1/4}. 
\label{eq:rcF}
\end{equation}
Since both radii have to be equivalent, the deceleration parameter of this system is determined to be $ q_\mathrm{s} = -3 $. 
This value brought a characteristic of viscous uniformity, 
corresponding to $H_\mathrm{s} \approx H_0/6$~\cite{Agatsuma_2020}. 

According to GR, binary star systems dominated by GW reaction have to be the slowest orbital decay. 
However, both PSR B1534+12~\cite{Stairs2002,B1534+12} and PSR B1913+16~\cite{Weisberg2010,Weisberg2016} show slower decay than the theory, called the excess $\dot{P_b}$, with total anomaly of about $2.5\,\sigma$~\cite{Agatsuma_2020}. 
A faster decay is possible by other energy losses, but slower decay is forbidden by GR. 
The above model can explain this deviation from GR. 
The cosmic drag picture is a promising model, featured by scale-independent and uniform expansion. 

This is still a hypothesis due to the consistency with few observations of the $2.5\,\sigma$ anomaly. 
In order to validate it furthermore, it is significant to scrutinize other observations. 
This paper shows additional two evidential binary pulsars with future test candidates. 

\section{Excess orbital decay}
The estimation of the orbital decay in a binary star system is written by
\begin{equation}
\dot{P_b}^\mathrm{obs} = \dot{P_b}^\mathrm{Kin} + \dot{P_b}^\mathrm{GW} + \dot{P_b}^\mathrm{X} .
\label{eq:OD}
\end{equation}
Here, $ \dot{P_b}^\mathrm{obs} $ is the observed orbital decay (time derivative of the orbital period), 
$ \dot{P_b}^\mathrm{Kin} $ is the all kinematic contributions in a galactic potential, 
$ \dot{P_b}^\mathrm{GW} $ is the theoretical estimate from GW reaction (intrinsic decay), 
$ \dot{P_b}^\mathrm{X} $ is the excess $\dot{P_b}$. 
Other contributions like the mass loss are omitted. 
The conventional model (no local expansion) expects $ \dot{P_b}^\mathrm{X} =0 $ 
when GW emission is a dominant factor of the energy loss. 

In the cosmic drag picture, the local cosmic expansion brake the orbital decay, 
which makes $ \dot{P_b}^\mathrm{X} $ as predicted below. 
According to \cite{Agatsuma_2020}, the orbital decay by GW emission is described by  
\begin{equation}
\frac{ \dot{P_b} }{ P_b }
= \frac{3}{2} \left[ 
H_\mathrm{s} - \frac{ K_\mathrm{gw} }{ r^{4} }  \frac{ 1+(73/24)e^2 + (37/96)e^4 }{ (1-e^{2} )^{7/2} }
\right] , 
\label{eq:dpdt}
\end{equation}
including the contribution from the cosmic drag. 
Here, $r$ is the orbital radius (as the eccentricity $ e=0 $) or the semimajor axis (as $e \neq 0$). 
Thus, the excess $ \dot{P_b} $ by the cosmic drag ($ \dot{P_b}^\mathrm{CD} $) is the first term of it: 
\begin{equation}
\dot{P_b}^\mathrm{CD} 
= \frac{3}{2} H_\mathrm{s} P_b. 
\label{eq:PbCD}
\end{equation}
This is applicable for $ r \leq r_\mathrm{c} $. 
Since the system can be regarded as the Keplerian orbit around $r_\mathrm{c}$, 
the orbital period is written by 
\begin{align}
P_b 
&= 2 \pi \left[ 
 \frac{ r^{3} }{ G (m_1 + m_2) }
\right]^{1/2} \\
&= 2 \pi \left[ 
 \frac{ k^{3} }{ G (m_1 + m_2) }
\right]^{1/2} \left( \frac{ K_\mathrm{gw} }{ H_\mathrm{s} } \right) ^{3/8} 
\label{eq:Pb}
\end{align}
where the radius is normalized by $ r_\mathrm{c} $ and $ k=r/r_\mathrm{c} $ is used. 
From the above equations, the expected excess is 
\begin{equation}
\dot{P_b}^\mathrm{CD} 
= 3 \pi H_\mathrm{s} \left[ 
 \frac{ k^{3} }{ G (m_1 + m_2) }
\right]^{1/2} \left( \frac{ K_\mathrm{gw} }{ H_\mathrm{s} } \right) ^{3/8} 
\label{eq:PbCD1}
\end{equation}
with viscous uniformity ($H_\mathrm{s} = H_0 / 6$)~\cite{Agatsuma_2020}. 

For a region of $ r_\mathrm{c} \leq r \ll R_\mathrm{c} $, 
as discussed in the prior paper~\cite{Agatsuma_2020}, 
the cosmic drag completely cancel the GW reaction (orbital decay), 
which indicates 
\begin{equation}
\dot{P_b}^\mathrm{CD} = -\dot{P_b}^\mathrm{GW}. 
\label{eq:PbCD2}
\end{equation}
Here, $ \dot{P_b}^\mathrm{GW} $ is obtained from Eq.\,(\ref{eq:dpdt}) as $ H_\mathrm{s} = 0 $. 
A fixed orbit is expected in spite of GW emission. 

If the observed $ \dot{P_b}^\mathrm{X} $ is consistent with the prediction $ \dot{P_b}^\mathrm{CD} $, 
it supports the cosmic drag picture. 

\section{The third and fourth evidential objects}
PSR J1012+5307 has become observable of the cosmic drag by recent refinements of parameters~\cite{Ding_2020}. 
J1012+5307 is a binary system, 
which consist of a pulsar of $1.82\,M_\odot$ and a white dwarf of $0.174\,M_\odot$ (NS-WD system) 
with a small eccentricity ($e=1.2 \times 10^{-6}$). 
The orbital radius slightly exceeds the critical radius ($r = 1.24 r_\mathrm{c}$) according to the orbital period $P_b$ of 0.6 days. 
It means that the local expansion cancels the orbital decay of GW emission under the cosmic drag picture, 
and hence, the orbit should asymptote Eq.\,(\ref{eq:impEOM1}) (almost Keplerian) without orbital decay~\cite{Agatsuma_2020}. 
When the system follows Keplerian orbit without orbital decay, 
$ \dot{P_b}^\mathrm{obs} = \dot{P_b}^\mathrm{Kin} $ should be observed in Eq.\,(\ref{eq:OD}). 
In other words, $ \dot{P_b}^\mathrm{X} = -\dot{P_b}^\mathrm{GW} $ (Eq.\,(\ref{eq:PbCD2})) is expected 
when $ \dot{P_b}^\mathrm{GW} $ is included in the estimate. 
According to the latest observation of PSR J1012+5307 (see Table 5 in \cite{Ding_2020}), 
the relevant parameters are 
\begin{eqnarray}
 \dot{P_b}^\mathrm{GW} &=& -13 \times 10^{-15} ,    \\
 \dot{P_b}^\mathrm{X} &=& (10.6 \pm 6.1) \times 10^{-15} , 
\end{eqnarray}
which is consistent with the above interpretation. 
This system looks a fixed orbit as $ \dot{P_b}^\mathrm{X} $ cancels $ \dot{P_b}^\mathrm{GW} $. 
It may be explained by the local cosmic expansion. 
This is the first hint to support the cosmic drag picture for both a region of $r>r_\mathrm{c}$ and an NS-WD binary system. 

Besides, a binary neutron star PSR J1906+0746 is also marginally evidential. 
The predicted $ \dot{P_b}^\mathrm{CD} $ is $ 0.8 \times 10^{-14} $, 
which is consistent with the observed $ \dot{P_b}^\mathrm{X} $ of $ (3\pm 3) \times 10^{-14} $ 
(see Table 3 in \cite{VanLeeuwen_2015}). 
The deviation from the conventional model is only $1\,\sigma $. 

By adding anomalies in the above two evidential objects to the previous value ($2.5\,\sigma $), 
the total significance beyond the conventional model is about $3.6\,\sigma $ (99.97\%). 
Their relevant parameters are listed in Table\,\ref{tbl:Evidence}. 
The observed $ \dot{P_b}^\mathrm{X} $ in J1012+5307, B1534+12, and J1906+0746 have a good agreement with the cosmic drag prediction ($ \dot{P_b}^\mathrm{X} \approx \dot{P_b}^\mathrm{CD} $), 
and B1913+16 has a close value. 
All four independent binaries in Table\,\ref{tbl:Evidence} have a positive $ \dot{P_b}^\mathrm{X} $ (slower decay) 
with $1.0 - 1.8\,\sigma $ significance, 
which is a forbidden region of GR without the cosmic drag. 

\section{Test region and candidates}
There is an observable range of the cosmic drag in the binary pulsars. 
For a region of $ r < r_\mathrm{c} $, the shorter the radius, the smaller the expansion effect. 
For a region of $ r > r_\mathrm{c} $, the expected difference ($ \dot{P_b}^\mathrm{CD} = -\dot{P_b}^\mathrm{GW} $) is getting smaller and smaller with increasing the radius. 
For the comparison, $ \dot{P_b}^\mathrm{CD} $ has to be larger than the observational uncertainty.  
The binary pulsars with too small separation or too large separation are out of scope. 

\begin{figure}[htb]
\includegraphics[width=8.5cm]{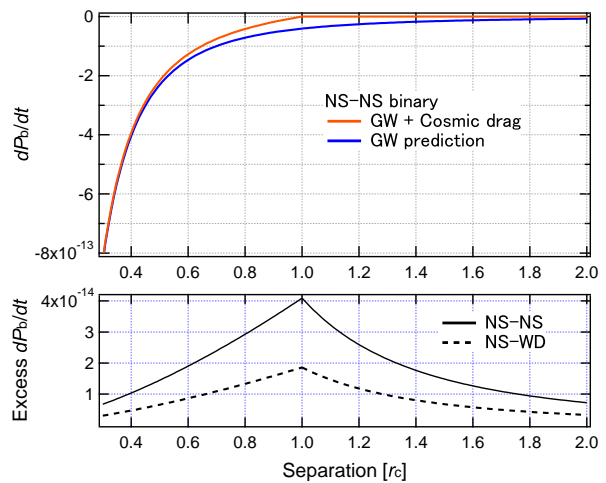}%
\caption{\label{fig:Difference_Pbdot}Rough estimates of $\dot{P_b}$ and the excess $\dot{P_b}$ due to the cosmic drag ($\dot{P_b}^\mathrm{CD}$). For simplicity, the same mass condition ($1.4\,M_\odot$) for neutron star (NS) and the 1/10 ratio ($0.14\,M_\odot$) for white dwarf (WD) are assumed with $e=0$. Viscous uniformity ($H_\mathrm{s} = H_0 / 6$) is adopted for $ r<r_\mathrm{c} $. 
} 
\end{figure}

Figure~\ref{fig:Difference_Pbdot} (top) shows $ \dot{P_b}^\mathrm{GW} $ with/without $ \dot{P_b}^\mathrm{CD} $ for NS-NS binary calculated by the above equations\,(\ref{eq:dpdt}), (\ref{eq:PbCD1}), and (\ref{eq:PbCD2}). 
Fig.~\ref{fig:Difference_Pbdot} (bottom) shows the expected $ \dot{P_b}^\mathrm{CD} $ for both NS-NS binary and NS-WD binary. 
Modern observations have an uncertainty of $ \dot{P_b}^\mathrm{X} $ of about $0.4-1.1 \times 10^{-14}$. 
Therefore, the observable range requires $ \dot{P_b}^\mathrm{CD} \gtrsim 1.0 \times 10^{-14}$, 
which is roughly $ 0.4\, r_\mathrm{c} \lesssim r \lesssim 1.8\, r_\mathrm{c} $ for NS-NS binary, 
corresponding to 
\begin{equation}
 0.2 \lesssim P_b \lesssim 2.0 \,\, \mathrm{[days]} 
\end{equation}
as $ P_b = 0.8 $ days at $ r=r_\mathrm{c} $. 
In the case of NS-WD binary, the observable range is shallower ($ 0.6\, r_\mathrm{c} \lesssim r \lesssim 1.4\, r_\mathrm{c} $), 
corresponding to 
\begin{equation}
0.2 \lesssim P_b \lesssim 0.6 \,\, \mathrm{[days]} 
\end{equation}
as $ P_b = 0.4 $ days at $ r=r_\mathrm{c} $. 
Note that $ H_\mathrm{s} $ is observable only for $ r < r_\mathrm{c} $ by Eq.\,(\ref{eq:PbCD}). 
Increase of mass and/or eccentricity expand the observable range because the increased $r_\mathrm{c}$ makes the peak excess higher. 
The He star companion, redbacks, and black-widow pulsars would be difficult for this test 
because the binary evolution is not GW dominant, meaning increased uncertainties. 

In addition to the evidential four binary pulsars, 
there are six test candidates, which can estimate $\dot{P_b}^\mathrm{CD}$ in this observable range. 
Their parameters are listed in Table\,\ref{tbl:Candidates}. 
In this list, the first two (J0509+3801 and J0751--1807) have no information about $ \dot{P_b}^\mathrm{X} $. 
The last two (J1829+2456 and B2127--11C) are uncertain values for $ \dot{P_b}^\mathrm{X} $. 
Except for B2127--11C, only J1738--0333 meets the conventional model ($ \dot{P_b}^\mathrm{X} = 0 $) in this region. 
J1756--2251 has a negative $ \dot{P_b}^\mathrm{X} $ (faster decay), 
which fit neither the conventional model nor cosmic drag picture. 

\begin{table*}
\begin{threeparttable}[htb]
\caption{\label{tbl:Evidence} Evidential binary pulsars.}
\begin{ruledtabular}
\begin{tabular}{c|c|c|c|c|c|c|c|c}
Binary pulsar & Pulsar mass & Companion mass  & Eccentricity & Orbital period & Separation 
& $\dot{P_b}^\mathrm{X}$  & $\dot{P_b}^\mathrm{CD}$ &Reference \\
& $m_1 (M_\odot)$ & $m_2 (M_\odot)$ & (e)\tnote{a} &   $P_b$ (days)\tnote{a}  &  $(r_\mathrm{c})$\tnote{b}   
& ($10^{-14}$) & ($10^{-14}$)\tnote{c}\, & \\
\hline   
J1012+5307 & 1.817\tnote{d} & 0.174(11)\tnote{d}  & $1.2 \times 10^{-6}$  & 0.605  & 1.24(3) & 1.1(6) & 1.3(1) & \cite{Ding_2020} \\             
B1534+12    & 1.3330(2)       & 1.3455(2)             & 0.274                     & 0.421  & 0.57  & 2.0(1.1)  & 2.1 & \cite{B1534+12} \\ 
J1906+0746 & 1.291(11)       & 1.322(11)             & 0.085                 & 0.166  & 0.35 & 3(3) & 0.8 & \cite{VanLeeuwen_2015} \\
B1913+16    & 1.438(1)        & 1.390(1)               & 0.617                     & 0.323  & 0.28 & 0.5(4),  0.8(5)  & 1.6 & \cite{Weisberg2016}, \cite{Weisberg2010} \\
 \end{tabular}
 \end{ruledtabular}
\end{threeparttable}
\end{table*}

\begin{table*}
\begin{threeparttable}[htb]
\caption{\label{tbl:Candidates} Test candidates.}
\begin{ruledtabular}
\begin{tabular}{c|c|c|c|c|c|c|c|c}
Binary pulsar & Pulsar mass & Companion mass  & Eccentricity & Orbital period & Separation 
& $\dot{P_b}^\mathrm{X}$ & $\dot{P_b}^\mathrm{CD}$ &Reference \\
& $m_1 (M_\odot)$ & $m_2 (M_\odot)$ & (e)\tnote{a} &   $P_b$ (days)\tnote{a}  &  $(r_\mathrm{c})$\tnote{b}   
& ($10^{-14}$) & ($10^{-14}$)\tnote{c}\, & \\
\hline   
J0509+3801  & 1.34(8)      & 1.46(8)   & 0.586                             & 0.380  & 0.34(1)   &  ---  & 1.9 & \cite{Lynch_2018} \\             
J0751--1807 & 1.64(15)\tnote{e} & 0.16(1)\tnote{e}   & $3.3 \times 10^{-6}$ & 0.263  &  0.74(2)  & ---  & 1.3 & \cite{Desvignes_2016}  \\ 
J1738--0333 & 1.46(6)     & 0.181(8) & $3.4 \times 10^{-7}$          & 0.355  & 0.89(1)   & 0.2(4)   & 1.7 & \cite{Freire_2012} \\
J1756--2251 & 1.341(7)   & 1.230(7)  & 0.181                             & 0.320  & 0.52  & $-1.7^{+0.9}_{-0.6}$  & 1.6  & \cite{Ferdman_2014} \\
J1829+2456  & 1.306(4)    & 1.299(4)  & 0.139                         & 1.176  & 1.25   & $-2.1(1.1)$\tnote{f} \,  & 2.3 & \cite{Haniewicz_2021} \\ 
B2127--11C   & 1.358(10)  & 1.354(10) & 0.681                         & 0.335  & 0.25    & $-1(13)$  & 1.6   & \cite{Jacoby_2006}  \\
 \end{tabular}
 \end{ruledtabular}
  \begin{tablenotes}
   \item[a] The number of digits are rounded for simplicity. 
   \item[b] Semimajor axis when $e \neq 0$. The uncertainty is due to mass estimation but it ($\sim 2\%$) from the Hubble tension is omitted. 
   \item[c] The main uncertainty is due to mass estimation for $r>r_\mathrm{c}$, or the Hubble tension of $\sim7\%$ for $r<r_\mathrm{c}$. The latter is omitted. 
   \item[d] The pulsar mass is derived from the companion mass~\cite{Antoniadis_2016} and mass ratio of 10.44~\cite{Sanchez_2020}. 
   \item[e] For the mass estimation, $ \dot{P_b}^\mathrm{GW} $ is used as the true value ($ \dot{P_b}^\mathrm{X} = 0 $).  
   \item[f] Not claimed value. This is derived from DDH solution but there is a discrepancy with DDGR solution.  
  \end{tablenotes}
\end{threeparttable}
\end{table*}

\vspace{4.5pc}
\section{Discussion}
\label{sec:discussion}
In the observable range of the cosmic drag, there are ten binary pulsars at least. 
At the present observations, four binaries of them have not reached the precision of interest. 
Four of the others (six) are almost consistent with the cosmic drag prediction. 
From their results, anomalies of the conventional (no expansion) model appear in the forbidden region with a total significance of about $3.6\,\sigma$. 
It is worth considering although the statistical significance has not reached a criteria of the ``discovery ($5\,\sigma $)''. 

As a possible reason of $ \dot{P_b}^\mathrm{X} $, 
a distance estimate by the dispersion measure is doubted in PSR B1534+12~\cite{B1534+12,Stairs2002}. 
While, the distance uncertainty cannot be a reason of $ \dot{P_b}^\mathrm{X} $ in PSR J1012+5307~\cite{Ding_2020}  
because this observation employs a parallax-based distance with less uncertainty. 

PSR J1738--0333 and PSR J1756--2251 have a discrepancy with the cosmic drag picture. 
Their results of $ \dot{P_b}^\mathrm{X} < \dot{P_b}^\mathrm{CD} $ (faster decay) are possible when uncounted energy losses contribute to $ \dot{P_b}^\mathrm{obs} $ (i.e. not GW dominant). 
If the cosmic drag picture is correct, J1738--0333 and J1756--2251 may be such situation or any systematic issue.  
It would be likely to reach $ \dot{P_b}^\mathrm{CD} $ observable by refining parameters as well as the case of J1012+5307~\cite{Ding_2020}. 

There is another interpretation of $\dot{P_b}^\mathrm{X}$ as alternative theory of gravity to change $G$ with time and/or gravitational dipole radiation (e.g. tested in~\cite{Lazaridis_2009,Ding_2020}). 
The main difference of the cosmic drag from them is the limited observable range around $ r_\mathrm{c} $ as shown in Fig.~\ref{fig:Difference_Pbdot}. 

\section{Acknowledgements}
I am grateful to Dr. Alberto Vecchio and Dr. Andreas Freise for their great support to continue this study. 
I thank Dr. Kazuhiro Yamamoto for a useful discussion. 
I appreciate Dr. Paulo Freire and his team for confirming measurement values in J1829+2456. 


\end{document}